\documentclass[Letter]{aa}
\usepackage{natbib}
\usepackage{graphicx}
\usepackage[varg]{txfonts}
\usepackage[colorlinks=true,citecolor=blue]{hyperref}
\begin{document}

   \title{Longitudinal drift of Tayler instability eigenmodes as a possible explanation
   for super-slowly rotating Ap stars}

   \author{L.\,L.~Kitchatinov
          \and
          I.\,S.~Potravnov
          \and
          A.\,A.~Nepomnyashchikh
          }

    \institute{Institute of Solar-Terrestrial Physics SB RAS, PO Box
291, Irkutsk 664033, Russia \\ \email{kit@iszf.irk.ru}
             }

   \date{Received 2 March 2020/Accepted 28 May 2020}
\abstract {Rotation periods inferred from the magnetic variability of
some Ap stars are incredibly long, exceeding ten years in some cases. An
explanation for such slow rotation is lacking.}
{This paper attempts to provide an explanation of the super-slow
rotation of the magnetic and thermal patterns of Ap stars in terms of
the longitudinal drift of the unstable disturbances of the kink-type
(Tayler) instability of their internal magnetic field.}
{The rates of drift and growth were computed for eigenmodes of Tayler
instability using stellar parameters estimated from a structure model
of an A star. The computations refer to the toroidal background magnetic
field of varied strength.}
{The non-axisymmetric unstable disturbances drift in a
counter-rotational direction in the co-rotating reference frame. The
drift rate increases with the strength of the background field. For a
field strength exceeding the (equipartition) value of equal Alfven and
rotational velocities, the drift rate approaches the proper rotation
rate of a star. The eigenmodes in an inertial frame show very slow
rotation in this case. Patterns of magnetic and thermal disturbances of
the slowly rotating eigenmodes are
also computed. }
{The counter-rotational drift of Tayler instability eigenmodes is a possible
explanation for the observed phenomenon of super-slowly rotating Ap stars.}


    \titlerunning{Drift rates of Tayler instability eigenmodes and Ap stars rotation}
    \authorrunning{L.\,L.~Kitchatinov et al.}

    \maketitle
\section{Introduction}
The kink-type instability of magnetized plasma pinches, commonly dubbed
the \lq Tayler instability' \citep{T73} in stellar physics, can be
related to the Ap stars magnetism. The instability has been considered
as a key ingredient of a hypothetical dynamo in the radiative envelopes
of the stars \citep{S02,ZBM07,RKE12} or only as an agent shaping their
surface magnetic pattern \citep{AR11mnras,AR11asna}.

The extensive literature on Tayler instability is mainly focused on the
stability criteria and growth rates of unstable disturbances as the most
significant characteristics of the instability \citep[see, e.g.][and
references therein]{GBT81,S99,B06,KR08,BU13,Gea19}. Apart from the
growth rates, eigenvalues of the linear stability problem include the
oscillation frequency. In the case of non-axisymmetric ($m = 1$) Tayler
instability, the oscillation means a longitudinal drift of the
instability pattern. The drift represents a longitudinal propagation of
a global wave. A brief discussion of the drift rates in \citet{RK10}
revealed their two characteristic properties. Firstly, the drift rates in
a co-rotating reference frame are negative, meaning a
counter-rotation migration of the instability pattern as whole. Secondly,
the drift rates depend significantly on the background field strength. The
rate is small compared to the rate of rotation when the background
magnetic field is smaller than the equipartition value where the Alfven
velocity equals the rotation velocity. When the field approaches and
then exceeds the equipartition strength, the drift rate increases
sharply and then saturates at a value close to the rate of rotation. In
the super-equipartition case, the instability pattern is close to
resting in the \lq laboratory frame' of a side observer.

If Tayler instability does indeed shape patterns of the surface
inhomogeneity of  Ap stars \citep{AR11asna}, then rotation rates inferred from their
magnetic or photometric variability can be extraordinarily small. Some
Ap stars are indeed observed to rotate very slowly. The observed
phenomenon of super-slowly rotating Ap stars (SSRAp) was recently
discussed by \citet{M19,M17} and \citet{Mea19}. These stars show
rotation periods in excess of one hundred days, and beyond ten years in some
cases \citep[cf. Fig.\,1 in][]{M19}. An explanation for the SSRAp
phenomenon is lacking.

The main motivation for this Letter is to draw attention to the possible
relation between the Tayler instability drift rates and the observed
rotation of Ap stars. To this aim, we  compute drift rates and the structure of eigenmodes for Tayler instability of a subsurface toroidal
magnetic field.

\section{Model}
The model used for the computations in this paper is identical to that of
\citet{KR08}. Therefore, we do not reproduce all the equations of the
model but describe the method, assumptions, and approximations used in
it.
\subsection{Model design}
The model concerns the linear stability of a magnetic field in the
radiation zone of a star rotating uniformly with an angular velocity
$\Omega$. The stability equations are formulated for standard spherical
coordinates $(r,\theta,\phi)$ with the axis of rotation as the polar
axis. The background magnetic field $\vec{B}$ is assumed to be
axisymmetric and purely toroidal:
\begin{equation}
    \vec{B} = \vec{e}_\phi\sqrt{4\pi\rho}\sin\theta\cos\theta\ r\,\Omega_\mathrm{A} ,
    \label{1}
\end{equation}
where $\vec{e}_\phi$ is the azimuthal unit vector and the Alfven angular
frequency $\Omega_\mathrm{A}$ is introduced for convenience. The
differential rotation, which can be present on the pre-main sequence
evolutionary stage because of non-uniform contraction, rapidly converts a
primordial field into an axisymmetric configuration with a dominant
toroidal component \citep{S99}. The equator-antisymmetric field of
Eq.\,(\ref{1}) is  what is expected from its winding by differential
rotation.

The background state of uniform rotation and magnetic field of
Eq.\,(\ref{1}) can be unstable to small disturbances. Radial
displacements in stellar radiation zones are opposed by buoyancy forces.
The displacements are assumed small compared to the local radius $r$.
Our stability analysis is therefore local in radius and assumes a
wave-type radial profile $\exp(\mathrm{i}kr)$, but it is global in
horizontal dimensions. The Boussinesq approximation with divergence-free
velocity disturbances is adopted. It is convenient to express
disturbances of the magnetic field ($\vec{b}$) and velocity ($\vec{v}$)
in terms of scalar potentials of their poloidal ($P$) and toroidal ($T$)
parts \citep[cf.][]{C61}:
\begin{eqnarray}
    \vec{b} &=& \vec{r}\times\vec{\nabla}\left(T_b/r\right)
        + \vec{\nabla}\times\left(\vec{r}\times\vec{\nabla}({P_b}/r)\right) ,
    \nonumber \\
    \vec{v} &=& \vec{r}\times\vec{\nabla}\left(T_v/r\right)
        + \vec{\nabla}\times\left(\vec{r}\times\vec{\nabla}({P_v}/r)\right).
    \label{2}
\end{eqnarray}
With this representation, isolines of the toroidal potentials on
spherical surfaces of constant $r$ represent the toroidal field lines
and the poloidal potential defines the radial field components. These
properties are used in the following section to map the structure of the
unstable eigenmodes.

Equation (\ref{2}) is substituted into linearised induction and motion
equations to give four equations of our stability analysis. The fifth
equation of the complete equation system describes the entropy
disturbances caused by the radial displacements \citep{KR08}.

The dominant mode of the Tayler instability for the background field
(\ref{1}) is the non-axisymmetric mode of the azimuthal wave number $m =
1$ \citep{GBT81}. The linear stability analysis reduces to the
eigenvalue problem. Time and longitude dependencies of non-axisymmetric
eigenmodes are therefore combined into the common exponential function
$\exp\left(\mathrm{i}m\phi - \mathrm{i}\omega t\right)$. The complex
eigenvalue,
\begin{equation}
    \omega = \mathrm{i}\gamma + w,
    \label{3}
\end{equation}
includes the growth rate $\gamma$ (decay rate if $\gamma$ is negative)
and frequency $w$. Any constant phase $\phi - wt = const$ of the
non-axisymmetric instability pattern drifts in longitude with the rate
$\dot{\phi} = w$. Frequency $w$ is therefore the rate of the
horizontally global wave propagation in longitude. The drift rate
depends on the reference frame. \citet{RK10} find that the
eigenmodes of Tayler instability drift in counter-rotation direction in
the co-rotating frame. The drift rates in an inertial frame are equal to
the angular velocity of the instability pattern that a side observer
could see. The computed drift rates in what follows therefore refer to
the inertial frame.

Finite diffusion can be important for Tayler instability. The equation
system includes the thermal ($\chi$) and magnetic ($\eta$) diffusivities
and viscosity ($\nu$) via the dimensionless parameters
\begin{equation}
    \epsilon_\chi = \frac{\chi N^2}{\Omega^3 r^2},\ \
    \epsilon_\eta = \frac{\eta N^2}{\Omega^3 r^2},\ \
    \epsilon_\nu = \frac{\nu N^2}{\Omega^3 r^2},
    \label{4}
\end{equation}
where $N$ is the buoyancy frequency. Another parameter including the
stellar structure characteristics is the normalised radial wavelength,
\begin{equation}
    \hat{\lambda} = \frac{N}{\Omega k r} .
    \label{5}
\end{equation}
In former computations, maximum growth rates were obtained for
$\hat{\lambda} \approx 0.1$ \citep{KR08}. Results of Sect.\,3 correspond
to this value. Parameters of Eqs.\,(\ref{4}) and (\ref{5}) should be
inferred from a stellar structure model.
\subsection{Stellar structure parameters}
The structure of an A-star of $2.5M_\sun$, with an initial hydrogen content $H =
0.71$, metallicity $Z=0.02,$ and an age of 0.5\,Gyr was computed with the
code {\scriptsize MESA} by \citet{Pea11}, version {\scriptsize
11532}\footnote{\url{http://mesa.sourceforge.net}}. The star has a
central convective core of about $0.07R$, a \lq convective skin' of
about $0.01R$ on the surface, and the radiation zone in between. The
radius $R$ of the star at the given age is $2.95R_\sun$. The star with
an effective temperature of about 8980\,K belongs to spectral type A2.

\begin{figure}
   \includegraphics[width=\hsize]{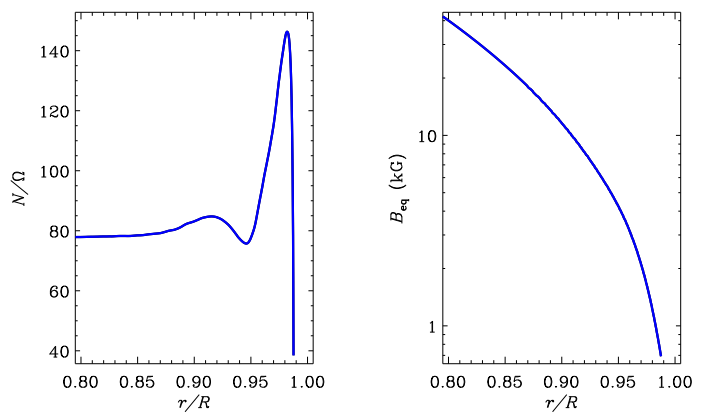}
   \caption{Normalised buoyancy frequency (left panel) and equipartition
field strength
        of Eq.\,(\ref{6}) (right panel) in the near-surface region of about 20\% of the star in radius. The plots correspond to the rotation period of 10 days.
            }
   \label{f1}
\end{figure}

The stabilising effect of the subadiabatic stratification is
characterised by the ratio $N/\Omega$ \citep{W81} present in the model
parameters of Eqs. (\ref{4}) and (\ref{5}). Figure\,\ref{f1} shows the
profile of this ratio for the characteristic rotation period
$P_\mathrm{rot} = 10$\,days of A-stars. The value of this ratio is close
to 80 in the upper part of the radiation zone. All estimations to follow
correspond to this value.  Results of the estimations differ
significantly between the cases when the Alfven angular frequency
$\Omega_\mathrm{A}$ is smaller or larger than the angular velocity
$\Omega$. The equipartition field amplitude
\begin{equation}
    B_\mathrm{eq} = \sqrt{\pi\rho}\,r\Omega,
    \label{6}
\end{equation}
of the equality $\Omega_\mathrm{A} = \Omega$ is also shown in Fig.\,\ref{f1}.

\begin{figure}
   \includegraphics[width=\hsize]{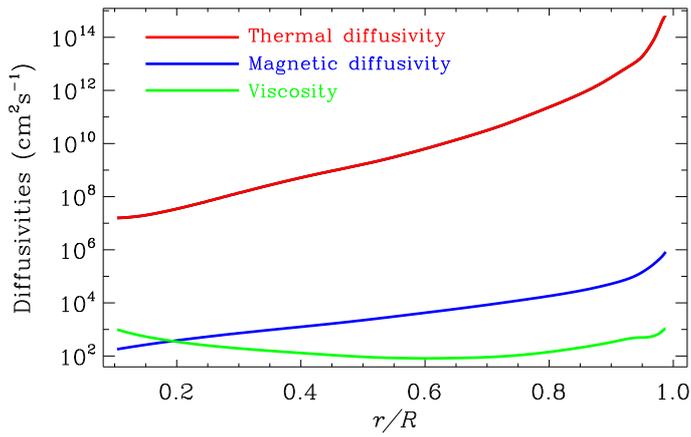}
   \caption{Radial profiles of the microscopic viscosity and thermal and
        magnetic diffusivities in the radiation zone of the star.
            }
   \label{f2}
\end{figure}

Figure\,\ref{f2} shows profiles of three basic diffusivities in the
radiation zone of the star. The magnetic diffusivity $\eta = 10^{13}
T^{-3/2}$\,cm$^2$\,s$^{-1}$ \citep[see e.g. Sect.\,3.2 in][]{BS05}
computed for the profile of temperature $T$ supplied by the stellar
structure model corresponds to the characteristic time of the Ohmic
decay ($\sim 10$\,Gyr) exceeding the age of the star. The results of the
following section refer to the steady background field of
Eq.\,(\ref{1}). Tayler instability is nevertheless sensitive to an
interplay between thermal and magnetic diffusion \citep{S99}.

Instability in the upper part of the radiation zone can be relevant to
surface observations. The diffusivity parameters of Eq.\,(\ref{4})
estimated for the radius $r = 0.9R$ read
\begin{equation}
    \epsilon_\chi = 8.0\times10^{-2},\ \
    \epsilon_\eta = 1.3\times10^{-9},\ \
    \epsilon_\nu  = 8.5\times10^{-12}.
    \label{7}
\end{equation}
Computations for the most interesting case of super-equipartition
background field ($\Omega_\mathrm{A} \geq \Omega$) can be performed with
this large spread in the diffusion parameters. However, the problem is
that computations for the subequipartition case of $\Omega_\mathrm{A} <
\Omega$ are not feasible with the large contrast between thermal and
magnetic diffusion of Eq.\,(\ref{7}). Only with the thermal diffusivity
reduced by about three orders of magnitude were computations for the
weak-field case possible. Results in the following section therefore refer to
the two cases of the diffusivities of Eq.\,(\ref{7}) for
$\Omega_\mathrm{A} \geq \Omega$ and to the reduced parameter of thermal
diffusion $\epsilon_\chi = 10^{-4}$ for a wider range of
$\Omega_\mathrm{A}$ including subequipartition fields.
\section{Results and Discussion}
\begin{figure}
   \includegraphics[width=\hsize]{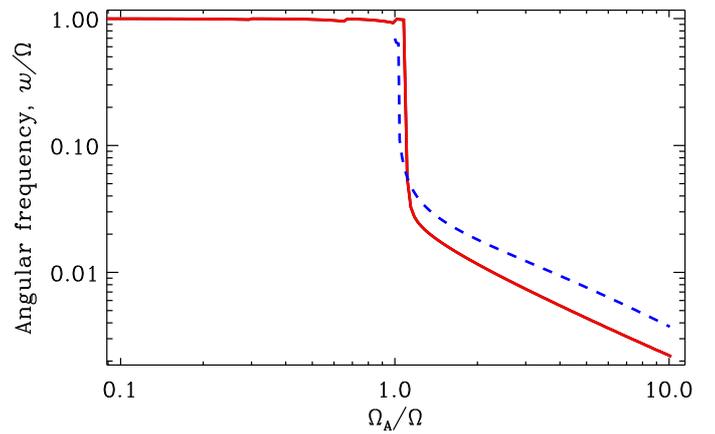}
   \caption{Drift rate of the most rapidly growing instability mode
        measured in units of the angular velocity of the star as a function
        of the normalised amplitude of the background field of Eq.\,(\ref{1}).
        The dashed line corresponds to the diffusion parameters of Eq.\,(\ref{7})
        and the full line corresponds to the reduced thermal diffusion of
        $\epsilon_\chi = 10^{-4}$.
            }
   \label{f3}
\end{figure}

\begin{figure}
   \includegraphics[width=\hsize]{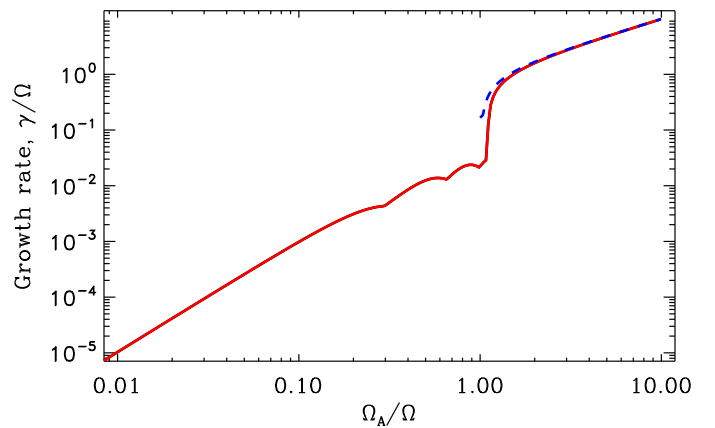}
   \caption{Fractional growth rate of the instability as a
        function of the normalised amplitude of the background field of Eq.\,(\ref{1}).
        The dashed line corresponds to the diffusion parameters of Eq.\,(\ref{7})
        and the full line corresponds to the reduced thermal diffusion of
        $\epsilon_\chi = 10^{-4}$.
            }
   \label{f4}
\end{figure}

Figure~\ref{f3} shows drift rates of the most rapidly growing (dominant)
instability mode in an inertial reference frame where the drift rates are
equal to the observable rate of rotation of the instability pattern. The
corresponding growth rates are shown in Fig.\,\ref{f4}. For the
subequipartition field strength of $\Omega_\mathrm{A}/\Omega < 1$, the
normalised drift rates are only marginally smaller than one, signifying
a co-rotation of the instability pattern with the star. The growth rates
for the subequipartition background field scale as $\gamma \propto
\Omega_\mathrm{A}^2/\Omega$ \citep{S99}. The scaling evidences a
stabilising effect of rotation on Tayler instability.

The most pronounced changes in Figs.\,\ref{f3} and \ref{f4} are
localised around the equipartition value of the field of
$\Omega_\mathrm{A} = \Omega$. The drift rate decreases and the growth
rate increases sharply with the field strength around this value.
Instability of the strong fields with $\Omega_\mathrm{A} > \Omega$ is
fast, its growth rate scales as $\gamma \propto \Omega_\mathrm{A}$. The
scaling shows that Tayler instability of strong fields is not sensitive
to rotation. Unstable disturbances of strong fields avoid the
stabilising effect of rotation by a counter-rotation drift. The drift
rates of Fig.\,\ref{f3} approach the power law
\begin{equation}
    w = c\,\Omega^2/\Omega_\mathrm{A}
    \label{8}
\end{equation}
for increasingly strong super-equipartition fields. The small
coefficient $c$ in this equation depends on the model parameters: $c
\simeq 0.037$ and $c \simeq 0.022$ for the dashed and full lines of
Fig.\,\ref{f3}, respectively.

The almost resting eigenmodes for super-equipartition fields is a robust
result; it was found for all considered parameter values and different
profiles of the background field \citep{RK10}. The counter-rotational
drift was also found by \citet{BU13mnras}. Oscillation frequencies in the
co-rotating frame of Figs. 1 to 3 by \citet{BU13mnras} are negative and
vary in direct proportion to the angular velocity for the
super-equipartition fields of $\Omega_\mathrm{A}
> \Omega$.

Two types of eigenmodes that differ in their equatorial symmetry can be
distinguished. The modes with equator-symmetric flow and entropy have an
equator-antisymmetric magnetic pattern. This combination is caused by the
anti-symmetry of the background field in  Eq.\,(\ref{1}). The other
symmetry type combines anti-symmetric flow and entropy with a symmetric
magnetic field. In our computations, modes of either symmetry type have
almost the same growth rate and the same drift. Figures~\ref{f3}
and \ref{f4} can be attributed to either symmetry type.

\begin{figure}
    \includegraphics[width=\hsize]{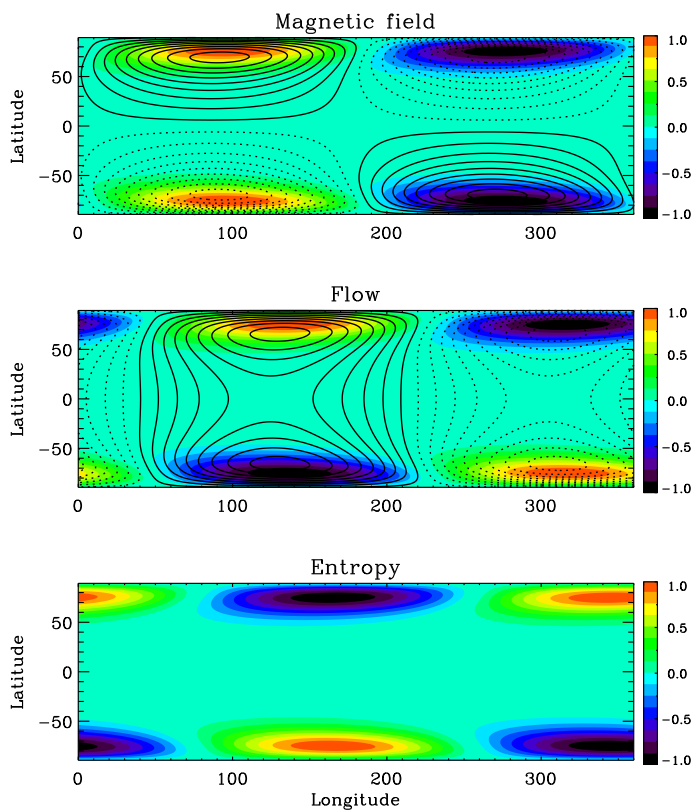}
    \caption{Structure of the most rapidly growing mode combining the
        equator-symmetric magnetic field with equator-antisymmetric velocity
        and entropy patterns, all for $\Omega_\mathrm{A}/\Omega = 1.5$.
        The top panel shows the magnetic field pattern: full (dashed) lines
        show the clockwise (anti-clockwise) circulation of the toroidal field
        vector and the colour scale indicates the radial component of the field.
        The middle panel shows a similar pattern for the velocity field.
        Thermal (entropy) disturbances are shown in the bottom panel.
        Colour scales are graduated in arbitrary units.}
    \label{f5}
\end{figure}

\begin{figure}
    \includegraphics[width=\hsize]{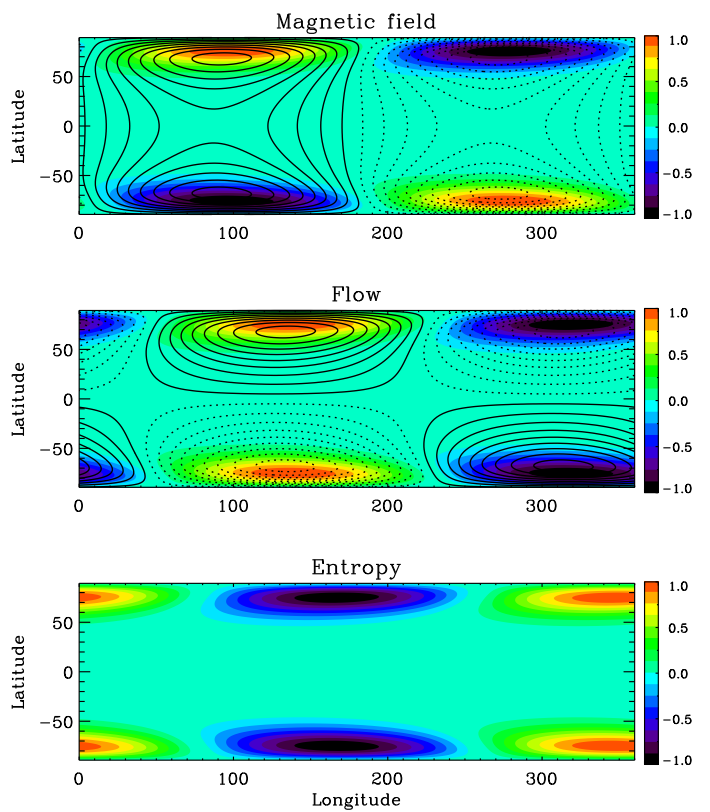}
    \caption{Same as in Fig.\,\ref{f5} but for the mode combining the
    equator-antisymmetric magnetic field with equator-symmetric flow and
    entropy.} \label{f6}
\end{figure}

The potentially observable patterns of the two types of equatorial
symmetry are shown in Figs.\,\ref{f5} and \ref{f6}. The reason for
coincidence of the corresponding drift and growth rates can be seen from
these figures. The eigenmodes of Figs.\,\ref{f5} and \ref{f6} are
concentrated to poles and they are small near the equator. Their
cross-equatorial link is therefore weak leading to practically
indistinguishable eigenvalues of the two symmetry types. The patterns of
these two symmetry types are very similar if viewed from a small
inclination angle to the rotation axes.

It may be noted that any superposition of the eigenmodes of
Figs.~\ref{f5} and \ref{f6} has the same rates of growth and drift. The
same is true for the patterns in Figs. 5 and 6 when shifted in longitude
by arbitrary phase. Any superposition of eigenmodes of different
symmetry or phase therefore gives  a mode with the same eigenvalue. The
superpositions make up patterns of different complexity that may explain
the diversity of observed magnetic patterns. In this context, an
interpretation by \citet{Mea16} of the unusual magnetic structure of
SSRAp star HD\,18078 in terms of a superposition of more simple magnetic
configurations can be quoted.

Comparison of the magnetic periods with the rotational periods
determined spectroscopically could be an observational test of our
hypothesis. However, the difficulty is that SSRAp stars usually possess
negligible spectroscopic signatures of rotation. The lines in their
spectra are narrow but split magnetically \citep{M19}. Due to
uncertainty in $\sin i$, this can  be interpreted  as either a result of
slow proper rotation of the star or its closeness to pole-on
orientation.  \citet{Bab60} suggested that the subgroup of the so-called
`sharp-lined Ap stars' is formed by stars observed under small
inclination angles. Such an interpretation does not contradict the
picture suggested by our model. Indeed, the polar magnetic patterns of
Figs.\,\ref{f5} and \ref{f6} are more easily observed with a pole-on
orientation. Rotational modulation of their magnetic zones remains
detectable with small but finite inclination angles.

A SSRAp star V1291 Aql (HD188041) can serve as an illustrative example.
A true rotational profile can be extracted from high-resolution spectra of
this star with sophisticated modeling techniques. \citet{Rom19}
determined the fundamental parameters of V1291 Aql using a spectrum
synthesised with a magnetic computational code. Their value for the
projected rotational velocity of the star's matter is $V\sin i =
2$\,km\,s$^{-1}$ and radius $R = 2.4R_\sun$. Even with a large
inclination angle of $i = 90\degr$, this leads to a proper rotation
period of about 60 days, which is about four times shorter than the 224 days
inferred by \citet{M19} from the star's magnetic variability. A small
inclination of $i\simeq10\degr$ leads to a \lq normal' value of
$P_\mathrm{rot} \simeq 10$\,days. This example is in favor of our
explanation of the SSRAp phenomenon.
\section{Conclusions}
The angular frequency of Tayler instability eigenmodes can be very small
(Fig.\,\ref{f3}). If the instability does indeed shape the surface
pattern of the magnetic field of the  Ap stars \citep{AR11asna,S02}, then
rotation periods inferred from magnetic (or photometric) variability can
exceed ten or even one hundred times the period of the stars' proper
rotation. The combined effect of the near pole-on orientation and the
azimuthal drift of the magnetic pattern due to Tayler instability could explain the observational phenomenon of SSRAp stars
\citep{M17,M19}.

Tayler instability has a low bound of several hundred Gauss for unstable
fields. The low bound could also be the cause of the magnetic dichotomy
found by \citet{Aea07}.

These explanations, if confirmed, point to Tayler instability as the
mechanism forming the magnetic fields of Ap stars.

The slow rotation of the instability pattern is reproduced for a
background field strength exceeding the equipartition value of
Eq.\,\ref{6}. The equipartition strength decreases with the proper
rotation period of a star. The equipartition field of Fig.\,\ref{f1} is
estimated for $P_\mathrm{rot} = 10$\,days, and should be reduced in
inverse proportion to $P_\mathrm{rot}$ for longer rotation periods. The
super-slow rotation of the instability pattern is therefore easier to
reproduce (with a weaker background field) in slower rotators.

There are several directions for improvement of our theoretical model
(waiving local approximation in radius, including nonlinearities, etc.).
Nevertheless, we believe that the results of our model can be offered
for a discussion prior to time-consuming advancements.
\begin{acknowledgements}
This work was supported by the Russian Foundation for Basic Research
(project 17-52-80064\_BRICS) and by budgetary funding of the Basic
Research programme II.16.
\end{acknowledgements}

\newpage

\bibliographystyle{aa}
\bibliography{paper}
\end{document}